\documentclass{hcj}
\usepackage{subcaption}
\journalyear{2014}
\journalvolume{1}
\journalissue{1}
\journalpages{101-131}
\articledoi{10.15346/hc.v1i1.2}
\journalcopyright{\copyright{} 2014, Pavlic \& Michelucci. CC-BY-3.0}

\title{Accurator: Nichesourcing for Cultural Heritage}
\author{
Chris Dijkshoorn\affil{Vrije Universiteit Amsterdam} \and
Victor de Boer\affil{Vrije Universiteit Amsterdam} \and
Lora Aroyo\affil{Vrije Universiteit Amsterdam} \and
Guus Schreiber\affil{Vrije Universiteit Amsterdam}
}
\authorrunning{C. Dijkshoorn, V. de Boer, L.M. Aroyo and A.T. Schreiber}

\begin{document}

\maketitle

\begin{abstract}
With more and more cultural heritage data being published online, their usefulness in this open context depends on the quality and diversity of descriptive metadata for collection objects. In many cases, existing metadata is not adequate for a variety of retrieval and research tasks and more specific annotations are necessary. However, eliciting such annotations is a challenge since it often requires domain-specific knowledge. Where crowdsourcing can be successfully used for eliciting simple annotations, identifying people with the required expertise might prove troublesome for tasks requiring more complex or domain-specific knowledge. Nichesourcing addresses this problem, by tapping into the expert knowledge available in niche communities.
This paper presents Accurator, a methodology for conducting nichesourcing campaigns for cultural heritage institutions, by addressing communities, organizing events and tailoring a web-based annotation tool to a domain of choice. The contribution of this paper is threefold: 1) a nichesourcing methodology, 2) an annotation tool for experts and 3) validation of the methodology and tool in three case studies. The three domains of the case studies are birds on art, bible prints and fashion images. We compare the quality and quantity of obtained annotations in the three case studies, showing that the nichesourcing methodology in combination with the image annotation tool can be used to collect high quality annotations in a variety of domains and annotation tasks. A user evaluation indicates the tool is suited and usable for domain specific annotation tasks.
\end{abstract}

\section{Introduction}
Many cultural heritage collections are currently being made available online~\citep{Mouromtsev2015, Szekely2013, DeBoer2012}. While such online collections can be valuable resources for the general public, scholars and professional users, their usefulness depends on correct and rich descriptions of the contained objects. Metadata describing objects is usually created by professionals working for the cultural heritage institution and typically meets the needs of other cultural heritage professionals. Many institutions lack the manpower to adapt data in order to better support different groups of users. Therefore, some institutions have turned to crowdsourcing, outsourcing tasks to a distributed and often anonymous group of people~\citep{Quinn2011}. For cultural heritage organizations, crowdsourcing proved to be a low cost solution to gather large quantities of descriptions~\citep{Chun2006, Ellis2012}.

While many institutions have gained significant experience with using crowdsourcing to collect large quantities of data, a remaining challenge is how to best harness the diversity in the crowd to solve difficult tasks in a sustainable fashion. Describing collection objects is a knowledge-intensive task, due to the variation in types of objects, diversity in subject matter and sometimes hidden symbolic meaning. Accurately annotating objects therefore often requires domain-specific knowledge. At the moment the required expertise is unavailable in an organization and when it is unfeasible to hire professionals to do the work, it is fruitful to reach out to experts within the crowd. 

Nichesourcing is a type of crowdsourcing, where groups of people with domain-specific knowledge are involved in the annotation process~\citep{DeBoer20122}. We call these groups of enthusiasts niche communities. There are numerous niche communities out there, revolving around lost of different domains. The advantages of nichesourcing are: 1) contributors are intrinsically motivated 2) potential of obtaining annotations of increased quality 3) knowledge-intensive annotation tasks can be executed. Where \citet{DeBoer20122} introduced the idea of nichesourcing and discussed small-scale case studies, a structured methodology was missing. We here present a repeatable and sustainable methodology as well as an open-source tool to support nichesourcing. We validate both using three extensive real-world case studies. The contribution of this paper is threefold:

\begin{itemize}
\item Accurator nichesourcing methodology (Section~\ref{sec:method}), which provides a step-by-step guide to designing and executing a nichesourcing campaign
\item Open source annotation tool (Section~\ref{sec:tool}) that supports the nichesourcing process
\item Validation of methodology and tool in three case studies in different domains(Section~\ref{sec:case_studies}) 
\end{itemize}

The results from the three case studies are presented in Section~\ref{sec:results} and the paper is concluded with a discussion.

\section{Related work}
\label{sec:related_work}
Human computation is a field in which the human ability to carry out computational tasks is leveraged to solve problems that can not yet be solved by computers alone~\citep{Quinn2011}. Crowdsourcing is part of the human computation field and regards tasks that are outsourced to a large group of people, often using the internet as an intermediary~\citep{Doan2011}. Crowdsourcing proved to be a good way to gather image annotations at scale~\citep{Ahn2004, Raddick2010}. This was recognized by the cultural heritage community and crowdsourcing has been used to annotate objects such as paintings, maps and videos~\citep{Ellis2012, Simon2011, Gligorov2011}. Gathered annotations  proved to be complimentary to annotations provided by cultural heritage professionals and thereby improved the accessibility of collections~\citep{Chun2006, Gligorov2013}. Crowdsourcing turned out to be a novel way of engaging the public as well~\citep{Ridge2013}.

%

Despite many successes, some crowdsourcing projects fail to live up to their expectations. Research has been conducted in classifying different types of crowdsourcing initiatives, to predict their success based on project characteristics~\citep{Noordegraaf2014}. \citet{Sarasua2015} discuss guidelines to how Semantic Web technology can support design and management of crowdsourcing projects, while ~\citet{Yadav2016} introduce guidelines for designing platforms hosting multiple projects. Methodology papers that outline steps to successfully run a crowdsourcing campaign are however scarce. In this paper we specify a nichesourcing methodology, contributing to the work available on crowdsourcing methodologies. More specifically, the methodology addresses crowdsourcing challenges such as solving knowledge-intensive tasks, involving experts, motivating contributors and assuring high quality contributions.

Knowledge-intensive tasks can not be solved by anyone in the crowd. There are different approaches to obtain high quality results, contributors can for example be matched with tasks~\citep{Cosley2007, Difallah2013}. \citet{Ahn2004} proposed theme rooms, clustering tasks by domain and leaving the choice to the contributor. \citet{Oosterman2016} invite experts from online communities to annotate objects in a specific domain. Another approach is to teach contributors how to solve difficult tasks using a game~\citep{Traub2014}.



\citet{Goto2016} optimize combinations of improvement tasks in crowdsourcing workflows, considering the average ability of workers, the variance in ability of workers and improvement difficulty.


\section{Accurator Nichesourcing Methodology}
\label{sec:method}
\begin{figure}
        \centering
                \includegraphics[width=0.7\textwidth]{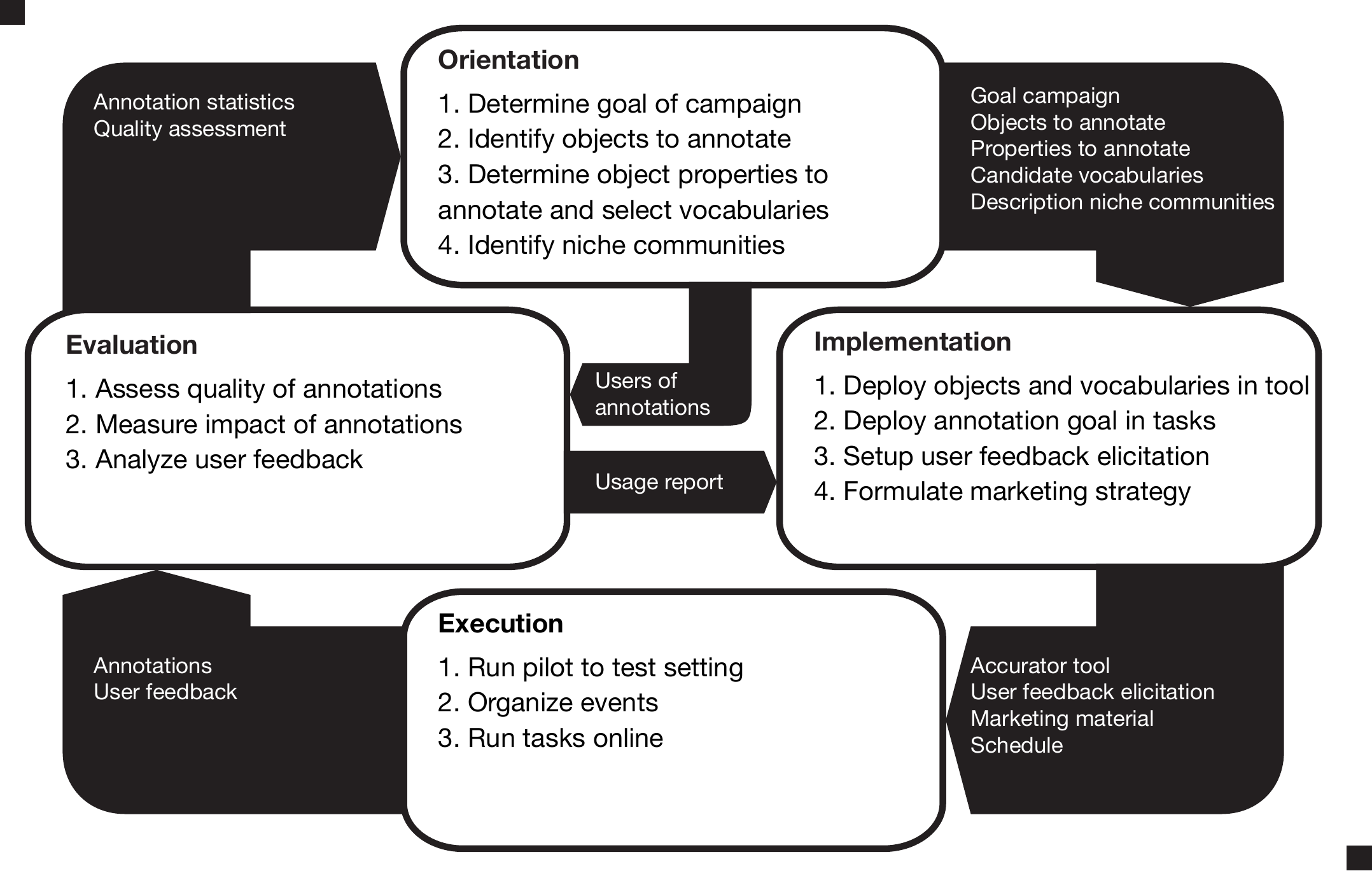}
        \caption{The four stages of the Accurator methodology.}
        \label{fig:method}
\end{figure}

Accurator is the name we use for our nichesourcing methodology as well as for the tool that supports running annotation tasks. In the next section, we describe the tool, whereas we here first describe the overall methodology. Figure~\ref{fig:method} provides a schematic overview of the methodology, which consists of four stages: orientation, implementation, execution and evaluation. These stages are further segmented into steps. In this section, we describe for each of these steps the input, output, action and challenges. We start by introducing the terminology used throughout this section.

The \textit{requester} initiates a nichesourcing campaign and corresponds often with the cultural heritage institution that owns the collection objects. The \textit{collection objects} are real world objects such as paintings, of which \textit{images} can be used in online applications. Images can be described with \textit{annotations}, which are often short textual descriptions or concepts. The \textit{task} combines a collection object with the sort of annotation requested. Tasks are solved by \textit{contributors}. We refrain from using the denomination worker, since there is no monetary reward given for solving tasks. The \textit{task difficulty} indicates how hard it is to solve a task and the \textit{contributor ability} is an indication of how well a contributor can solve hard tasks.
To enable solving tasks with a high task difficulty, nichesourcing campaigns involve \textit{niche communities}, groups of people possessing expert knowledge about a domain. A \textit{domain} is a area of interest, for example ornithology or fashion.

\newenvironment{method}{
\newcommand{\name}[2]{{\sffamily##1 ##2}\vspace{0.1cm}\\}
\newcommand{\inputoutput}[2]{\textit{input:} ##1 \textit{output:} ##2\vspace{0.1cm}\\}
\newcommand{\action}[1]{\textit{action:} ##1\vspace{0.1cm}\\}
\newcommand{\challenges}[1]{\textit{challenges:} ##1}
}{}

\subsection{Orientation stage}
In the first stage, the goal of the campaign is determined and the objects that need to be annotated to reach this goal are identified. Based on the characteristics of these objects, the required enrichment is determined, which guides the identification of niche communities who can provide such information.

\begin{method}
\name{1.}{Determine goal of campaign}
\inputoutput{annotation statistics}{goal campaign, users of annotations}
\action{The goal states what an institution wants to achieve with the annotations obtained during a nichesourcing campaign. The formulated goal can range from general (e.g. to improve access to a collection) to specific (e.g. to answer a digital humanities research question). A goal is based either on needs internal or external to the institution, therefore it is accompanied by an overview of the intended users of the annotations. A list of users provides clarity about who benefits from the data and gives an indication of who will take care of the collected information when the campaign is completed. At the end of the campaign, the goal is used to verify whether the gathered annotations have the desired impact. If the annotation statistics indicate the goal is reached, a new goal is formulated, otherwise a subsequent improved campaign is used to reach the current goal.}
\challenges{The goal has to be formulated in such a way that contributors deem it worthy to invest time into, fitting with their domain of interest. Additionally, it helps to validate the results of a campaign if it is possible to measure whether a goal is reached. For example, if the aim is to improve access to a collection, this can be measured by standard information retrieval metrics such as precision and recall.}
\end{method}

\begin{method}
\name{2.}{Identify objects to annotate}
\inputoutput{goal campaign}{objects to annotate}
\action{During this step a subset of objects is identified, which when correctly annotated will bring us closer to the formulated goal. This can be a) on the basis of automatic (data-driven) analysis; b) through manual selection of objects that require improvement or c) by analyzing user interactions with the collection. Most cultural heritage collections consist of objects that relate to a range of different domains. To be suitable for nichesourcing, the selected objects should share a domain, which can be matched with a community of experts.}
\challenges{Once a set of collection objects is either automatically or manually identified, preparation steps might be needed to ensure that basic metadata and a digital representation are available for every object in the set. Additionally, intellectual property rights should allow digital representations of the objects to be used online.}
\end{method}

\begin{method}
\name{3.}{Determine object properties to annotate and select vocabularies}
\inputoutput{objects to annotate}{properties to annotate and candidate vocabularies} 
\action{The object properties that need to be annotated to achieve the goal are identified during this step. More specifically, we distinguish properties of the objects that can be better described using numerical values, textual descriptions or concepts from structured vocabularies. One specific goal here is to identify structured vocabularies that can be used as values for the annotations. These vocabularies can be provided as input to the Accurator tool, which presents concepts of the vocabulary as options to the contributors.}
\challenges{Cultural heritage institutions have to carefully consider which vocabularies to use for describing collection objects. The suitability of vocabularies should be assessed in terms of completeness, accuracy and original context. It can, for example, be that the vocabulary was intended to be used in a completely different context and therefore does not contain the desired concepts, or that concepts represent a world view which is different from the institution. A lack of available labels in some language can pose a more direct problem.}
\end{method}

\begin{method}
\name{4.}{Identify niche communities}
\inputoutput{quality assessment}{description niche communities}
\action{To assess the feasibility of nichesourcing, the shared domain of the set of objects should match with a niche community. These communities are identified by contemplating on which people have the expertise to annotate the objects. The characteristics of an object can for example match with professionals outside the cultural heritage sector, or with hobbyists focusing on a certain topic. A common feature of niche communities is that they can be divided into even more specialized sub-niches. It is useful to identify such sub-niches, since later in the process it helps to assign tasks to contributors most knowledgeable of a sub-niche.}
\challenges{The description of niche communities should include ways of reaching out to the community, which is important for the marketing strategy in the implementation stage. Furthermore, the niche community should not only be determined on the basis of the match with the objects, but more importantly on the match with the missing information. It is not always straightforward to identify niche communities that match the selected objects and requested information. It is therefore important to allow interplay between the steps, adapting the selection or requested information to the communities available.}
\end{method}

\subsection{Implementation stage}
In this stage, the tool is deployed and tasks are designed that help reach the goal of the campaign. A marketing strategy is formulated to address the niche communities.

\begin{method}
\name{1.}{Deploy objects and vocabularies in tool}
\inputoutput{objects to annotate, candidate vocabularies}{accurator tool}
\action{In this step, the Accurator nichesourcing tool is deployed and relevant data is loaded. We describe the Accurator tool in more detail in Section~\ref{sec:tool}, but in general, this requires a requester to a) setup a server environment; b) install the nichesourcing tool; c) adapt the tool to the domain. Once deployed, data regarding the selected objects and vocabularies is loaded. A single instance of a tool can accommodate multiple campaigns, refraining an institution from having to deploy a tool for each iteration of the methodology.}
\challenges{Deploying the Accurator tool requires technical knowledge as well as appropriate infrastructure. Not every institution will have both readily available and therefore some might choose to outsource this step. Alternatively, an institution can choose to use existing online crowdsourcing platforms (e.g. Amazon Turk), thereby bypassing this problem. This has the downside that these platforms are not easily customized to support a particular domain.}
\end{method}

\begin{method}
\name{2.}{Deploy annotation goal in tasks}
\inputoutput{goal campaign}{accurator tool}
\action{During this step, the goal of the campaign is translated into smaller annotation tasks. Tasks combine objects with explicit requests for information and instructions of how this information should be provided. For a photograph, the requested information could be depicted persons, accompanied by the instruction to enter names into a text field. The identified structured vocabularies are related to requests, allowing rendering of suggestions for values to enter. Tasks are defined in the annotation tool through relating the identified objects to input fields, each accompanied by the information request and structured vocabulary.}
\challenges{The request for information and instructions have to be concise and unambiguous. If there is room for interpretation, this will have a negative impact on the consistency of the provided annotations. The concepts suggested can help normalize the input, but should fit the type of information requested.}
\end{method}

\begin{method}
\name{3.}{Setup user feedback elicitation}
\inputoutput{-}{user feedback elicitation}
\action{To get insights into the behavior of users and collect feedback, user elicitation mechanisms are setup. These mechanisms can be automated and unobtrusive, such as logging interactions with the annotation tool. An institution can also choose for more direct inquiring, for example by using questionnaires. Information gathered using these mechanisms is used to refine the orientation stage and can indicate the effectiveness of a marketing strategy. Furthermore, created user profiles can serve as input for automated quality assessment of annotations~\citep{Ceolin2012}.}
\challenges{Nichesourcing relies on the intrinsic motivation of contributors. To not annoy contributors and distract them from solving tasks, the elicitation mechanisms should be as unobtrusive as possible.}
\end{method}

\begin{method}
\name{4.}{Formulate marketing strategy}
\inputoutput{description niche communities}{marketing material and schedule}
\action{A marketing strategy is formulated to engage niche communities and capture the attention of contributors. This strategy includes a schedule that details when and how messages are communicated. Different outlets can be used, such as social media, newsletters and flyers. The choice of outlet depends on how the targeted niche community can best be reached. First communications are focussed on drawing attention to the campaign, by inviting people to participate in annotation events. Following an event, a message can be sent about the progress made, in addition to an invitation to keep contributing online. Subsequent communications are meant to entice people to keep participating in the campaign. At the end of the campaign, the impact of the annotations is emphasized, alongside pointing contributors towards new campaigns when possible.}
\challenges{It can be challenging to reach the niche communities identified during the orientation stage. Sometimes organizations that already rally events around the domain of interest can serve as a point of entry. These organizations are often different from the cultural heritage organization that owns the collection. Finding a niche representative within such an organization, who is willing to collaborate, greatly eases addressing potential contributors. Another strategy is to market the nichesourcing campaign together with a broader event associated with the domain, for example a National Week of Fashion or an exhibition organized by the institution. This allows institutions to combine the effort needed for marketing.}
\end{method}

\subsection{Execution stage}
With the tool deployed and the marketing strategy in place, the nichesourcing tasks can be executed. But first tasks deployed in the annotation tool are tested during a pilot.

\begin{method}
\name{1.}{Run pilot to test the setting}
\inputoutput{accurator tool}{-}
\action{To test the annotation tool and formulated tasks, a pilot is run with a limited number of members of the targeted niche community. During the pilot, issues are identified that should be addressed before the event. Depending on the type of issue, the subset of objects, selected vocabularies and tasks are refined.}
\challenges{For each issue an assessment has to be made whether it will apply to most members of the community and therefore warrants a follow-up action.}
\end{method}

\begin{method}
\name{2.}{Organize events}
\inputoutput{accurator tool, user feedback elicitation, marketing material, schedule}{annotations, user feedback}
\action{Organizing an annotation event is an essential element of an Accurator nichesourcing campaign. Besides being the first source of annotations and feedback, the event is used to engage the niche community. The organization of events constitutes of three aspects: timing, location and program. With respect to timing, enough time is needed to implement the marketing strategy and advertise the event in the niche community. To make the event as attractive as possible, the event should preferably take place at a location relevant to the domain of interest. This could be at the institutions of the collection owner, or at another place relevant to the domain. The program of the event includes an introduction and demonstration of the nichesourcing tool. After this, contributors use the tool to annotate the collection objects.   The event is concluded with a discussion, resulting in feedback which can be used during the evaluation stage. Optionally, the program can be extended with activities, functioning incentive for experts to participate.}
\challenges{It can be challenging to strike the right balance between time for annotating, discussion and extra activities. Enough time has to be available for annotating collection objects, in order to collect sizable amounts of annotations and to make sure that contributors have enough time to work with the tool to be able to provide feedback.}
\end{method}

\begin{method}
\name{3.}{Run tasks online}
\inputoutput{accurator tool, user feedback elicitation, marketing material, schedule}{annotations, user feedback}
\action{After the annotation event, the campaign continues online. Running the nichesourcing tasks online regards advertising the annotation tool and providing support to contributors. The interest sparked up by the event serves as initial input for advertising the tool. Updating contributors on the results of the annotation event helps to incentivize people to return at a later point in time and continuously add annotations to the collection. To sustain this attention and reach new contributors, the tool is advertised as outlined in the marketing campaign. Finding additional experts could be automated using techniques such as the one proposed by \citet{Oosterman2016}. In order for contributors to not get discouraged when they run into problems, adequate support has to be available.}
\challenges{To sustain the interest of contributors, a cultural heritage institution will have to invest in the support and marketing of the annotation tool. When a group of contributors is actively involved in the nichesourcing campaign, the effort of marketing and providing support can be shifted towards the community~\citep{Bevan2014}.}
\end{method}

\subsection{Evaluation stage}
At the end of the nichesourcing campaign, the impact and quality of the annotations is assessed. Feedback gathered during the campaign is used to improve subsequent campaigns.

\begin{method}
\name{1.}{Assess quality of annotations}
\inputoutput{annotations}{quality assessment}
\action{The quality of annotations is assessed during this step. Quality verification procedures can be manual processes or automated processes. Both can be used within a nichesourcing campaign, although their suitability should be assessed up front. An example of a manual process is reviewing (parts of) the annotations, by contributors or professionals. An example of an automated procedure is majority voting, in which the annotation is used that most contributors added to an object. An institution decides based on the assessment, to reject or improve annotations~\citep{Goto2016}. Institutions should consider publishing the annotations along with their quality assessment since further analysis of measured disagreement leads to new insights in crowdsourced data~\citep{Inel2014}.}
\challenges{A relatively naive method such as majority voting might be less appropriate for nichesourcing, since a small number of experts might be knowledgeable enough to provide a correct answer. An answer which might, in turn, contradict answers of other contributors. Other automated approaches would, therefore, be more suitable, for example considering trust in a contributor based on earlier annotations~\citep{Ceolin2012}.}
\end{method}

\begin{method}
\name{2.}{Measure impact of annotations}
\inputoutput{users of annotations, annotations}{annotation statistics}
\action{During this step, the verified annotations are deployed to investigate whether the goal is reached. If the goal is to improve accessibility and the user of the annotations is the institution, this for example entails exporting the data from the tool and incorporating the results into the collection data. At that point, a comparison of search performance of the collection with and without the annotations can provide an indication of impact~\citep{Gligorov2013}. If the goal cannot be reached, this evaluation serves as input for improving the next nichesourcing campaign, by for example adapting the set of objects or the properties to annotate.}
\challenges{Quantifying the impact of annotations can be difficult and depends on the formulated goal. Thereafter, it can be challenging to translate this evaluation towards adaptions of the next nichesourcing campaign.}
\end{method}

\begin{method}
\name{3.}{Analyze user feedback}
\inputoutput{user feedback}{usage report}
\action{Feedback is gathered during events as well as online. User feedback follows from sources such as questionnaires, discussions, support requests and interaction logs. Analyzing these sources can help to improve subsequent nichesourcing campaigns. Common feedback topics regard task complexity and appropriateness of the tool. If tasks are deemed too complex, changes can be made to the selection of objects, chosen properties to annotate and the niche community which is addressed. When tasks are too easy, other crowdsourcing approaches could be considered. Feedback regarding the tool can be addressed by improving the code or choosing a different platform to deploy tasks.}
\challenges{Operationalizing the gathered feedback, by improving new campaigns can be a challenge. It is, however, important to acknowledge feedback and improve the process. A contributor providing feedback took the time to work with the tool and provide feedback. If this feedback is taken seriously, a contributor might feel more inclined to contribute to a new campaign. Addressing problems with tooling requires technical skills which might not be available within an institution. The shortcomings of a tool could therefore be communicated to the contributors providing feedback, or programmers could be contacted to improve the tool. The Accurator annotation tool, which we discuss in the next section is open source, allowing anyone to improve the code as desired.}
\end{method}

\section{Accurator annotation tool}
\label{sec:tool}
To support the nichesourcing methodology, we present a tool called \textit{Accurator}. Accurator is an web-based annotation tool which can be instantiated for specific nichesourcing campaigns,
to allow contributors to annotate images of cultural heritage objects that are automatically assigned to them. This section describes the tool and more specifically its adaptability to a specific campaign and chosen domain, as well as usability design considerations. We conclude this section by discussing how the collected data can be used by other systems and how the annotations directly impact the search functionality of the tool.

The implementation of the tool is based on Semantic Web technology~\citep{Shadbolt2006}. The Accurator annotation tool is available as a package for the Cliopatria Semantic Web infrastructure~\citep{Wielemaker2016}. The back-end is written in the Prolog programming language, which facilitates direct access to the data layer~\citep{Wielemaker2007}. The front-end uses jQuery\footnote{\url{https://jquery.com}} and Twitter Bootstrap\footnote{http://getbootstrap.com} so contributors experience an interactive and responsive tool. The source code of the package is published online, along with an in-depth guide to how new instances can be deployed\footnote{\url{https://github.com/rasvaan/accurator}}.

\subsection{Adaptability to domain}
Accurator can be customized to fit a domain, by using config files containing \textit{domain definitions}. The definitions define links to 1) a specification of the annotation fields relevant to the domain, 2) elements of the interface tailored to the domain and 3) information that makes task assignment possible. Here we discuss annotation fields and interface adaption, task assignment follows in a separate subsection. 

\begin{figure}
	\centering
	\begin{subfigure}[b]{0.49\textwidth}
		\includegraphics[width=\textwidth]{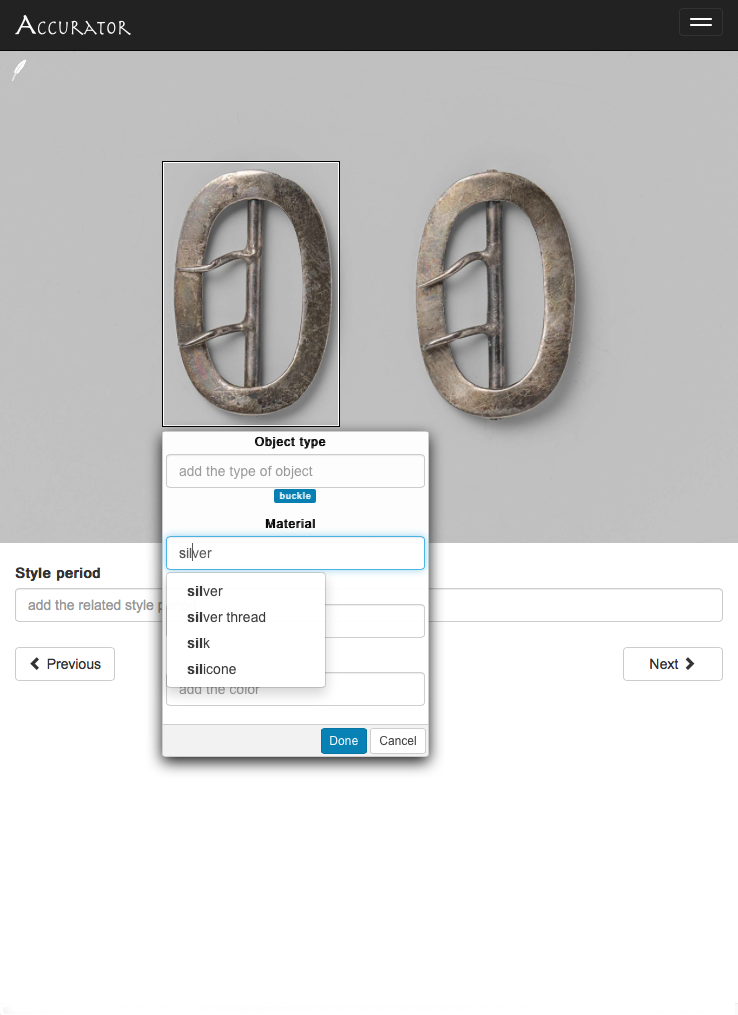}
		\caption{Fashion domain}
		\label{fig:interface_fashion_domain}
	\end{subfigure}
	\begin{subfigure}[b]{0.49\textwidth}
		\includegraphics[width=\textwidth]{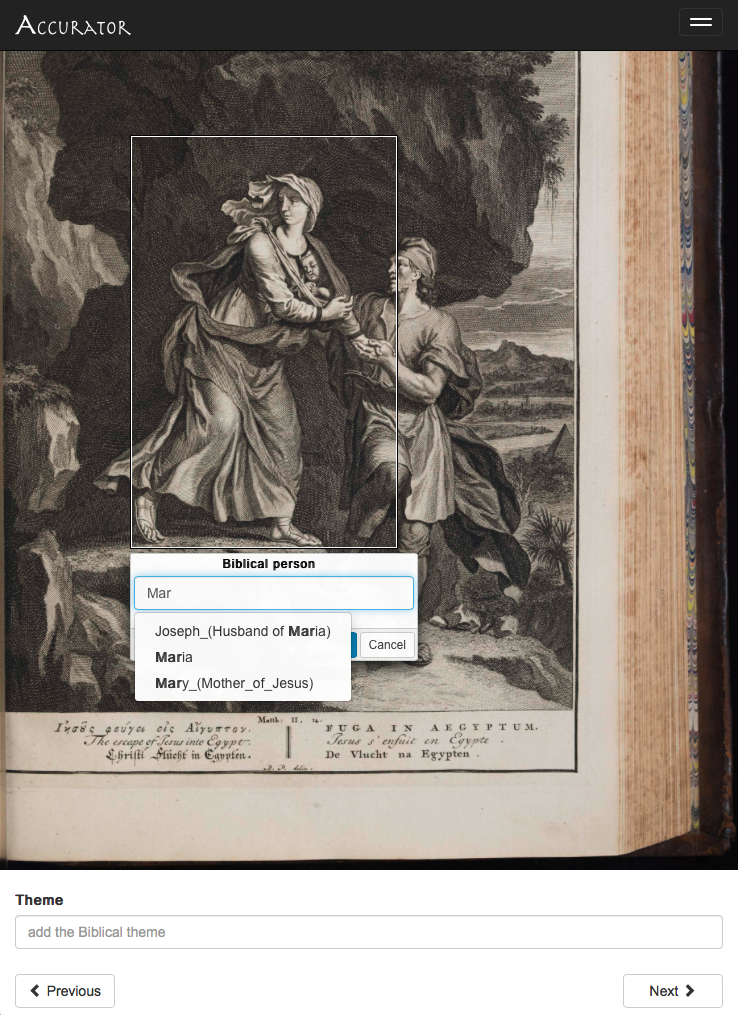}
		\caption{Bible domain}
		\label{fig:interface_bible_domain}
	\end{subfigure}
	\caption{Annotation interface of the Accurator tool, showing the fields that can be used to annotate objects in the fashion domain and bible domain.}\label{fig:interface}
\end{figure}

Annotation tasks are adaptable to the domain, a requester can specify field definitions for each of the annotation fields. These specifications include the field name, a short instruction and the type of field. Different types include radio buttons, check boxes and text fields. Text fields can use the autocompletion functionality, where a contributor starts to type and a drop-down menu renders alternatives related to this input, as shown in Figure~\ref{fig:interface}. The contributor can either choose to annotate the object using one of these alternatives or use the entered text. These alternatives originate either from a list of values added to the field definition or from a subset of a structured vocabulary. Accurator includes Prolog predicates to identify such subsets of vocabularies, for example based on a branch within a taxonomy.

Annotation fields can be defined as being about the object as a whole, or be defined as being about a specific part of the object. In the first case, annotation fields are presented to a contributor alongside the image. An example of this is the style period of fashion objects. In the second case, users can draw a bounding box in the image to identify the specific part of the object that the annotation concerns, as shown in Figure~\ref{fig:interface}. This allows users to annotate multiple specific elements of an object, for example, two birds of different species depicted on a print.

The default visual elements and text of the tool can be adapted as well, the default tagline used on the intro page of the tool ``Help us add information to artworks'' can, for example, be changed to one tailored to the fashion domain (e.g. ``Help us describe fashion''). At the same time, it is possible to add images, which brand the tool with visuals related to the niche.

\subsection{Task assignment}
\label{sec:task_assignment}
Task assignment concerns the matching of contributors with tasks. Accurator provides
three modes of task assignment: \textit{ranked}, \textit{sub-domain based} and \textit{recommendation}. Ranked is the default mode, which first filters out the objects already annotated by the user and sorts the remaining objects based on the total number of users that annotated them. A list of objects randomly picked from the least annotated objects is presented to the contributor. This is the default setting since it ensures a rapid increase in annotated objects.

In sub-domain based mode, contributors can choose in which sub-domain they would like to annotate. To this end, a hierarchy of general and more specific domains is created, by adding references in the config file of a domain definition. The fashion domain described in Section~\ref{sec:fashion} can, for example, be split into more specific domains such as costumes and jewelry. The availability of sub-domains triggers a finer grained mode of task assignment, as the objects presented to contributors are filtered based on the domain they belong to. The objects from the domain chosen by the contributor are then ranked according to the ranked method described above.

The third mode of task assignment is recommendation. Recommending suitable tasks to contributors might make the annotation process more accurate and efficient. With the Accurator tool, we experimented with recommendation based on the elicitation of expertise levels of contributors. To do this, a requester needs to decide on a list of expertise topics, elicit the expertise levels from the contributors and use this as input for a recommender algorithm. The list of topics is based on a structured vocabulary, referenced in the config file. In case of the bird domain, an example of topics can be a branch of the biological taxonomy. Contributors are asked to assess their expertise regarding each selected topic. The highest ranking topics are used as input for an explorative search algorithm, which uses the graph structure to find objects that are related to the expertise of the contributor~\citep{Wielemaker2008}. In Section~\ref{sec:results}, we evaluate the three different task assignment approaches and consider the feedback of the contributors.

\subsection{Usability}
Usability is important for crowdsourcing tools, and we argue that this is especially true for tools that are used for nichesourcing since nichesourcing relies on the intrinsic motivation of contributors. Wasting their goodwill because the tool is hard to use might make a requester miss out on valuable input. While the tool uses many Semantic Web techniques, as outlined in~\citep{Sarasua2015} we should not expect our contributors to be Semantic Web experts. The interface, therefore, hides technical aspects such as the persistent identifiers from contributors and uses textual labels of concepts and properties wherever available.

Part of the usability is presenting a tool in the language of the contributor. The primary language of the annotation tool is English, but many of the contributors prefer a different language. The tool supports translating textual elements of the interface, in a similar fashion as adapting texts to the domain. We translated the interface to Dutch, thereby customizing the usage for contributors from the Netherlands. The autocompletion alternatives are based on the labels of concepts of structured vocabularies. Oftentimes these labels are available in multiple languages. The tool is designed to render alternatives in the language of choice if available, otherwise falling back on English labels.

The Accurator tool is designed to work with all regular browsers, even older versions. Therefore most contributors will be able to use the tool on their own system. The registration procedure is simple and requires minimal information to be entered by potential contributors. Questions requesting additional information about contributors used for scientific purposes are spread out over multiple blocks, each appearing after a contributor added a specific number of annotations. Additionally, system administrators are advised to use simple domain names for the online tool, so contributors can easily remember how to access the instance. We evaluated how contributors perceive the usability of the tool using a questionnaire, the results are discussed in Section~\ref{sec:results}.

\subsection{Direct impact annotations}
Annotations added by contributors can be directly used by other systems and have a direct impact on the semantic search functionality of the Accurator tool. Annotation data is stored in the RDF triple store of the annotation tool, which is separated from the collection management system or catalogue of the institution. Using this architecture, systems that cope well with crowdsourced data can have direct access to new information, while systems relying on verified data can use exports of the information of which the quality is assessed. Storing the data using the Resource Description Framework (RDF)\footnote{\url{https://www.w3.org/RDF/}} and standardized data models improves the reusability of data. We continue by discussing the data model used within the tool, followed by a discussion of the impact of annotations on search and the ways of how the collected data can be made available.

\begin{figure}
       \centering
                \includegraphics[width=0.95\textwidth]{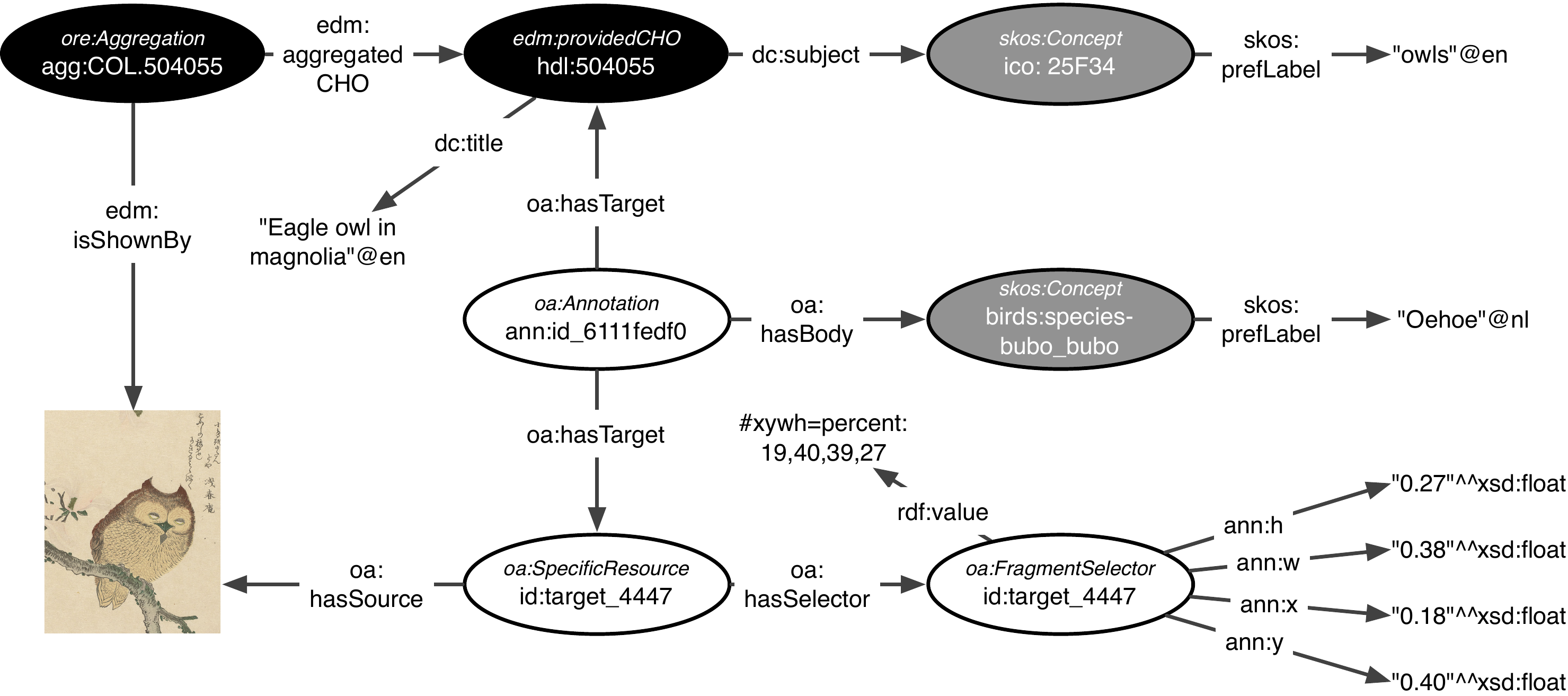}
        \caption{A graph representation of the print ``Eagle owl in magnolia'' annotated with the species of the depicted bird.}
        \label{fig:data_model}
\end{figure}

A graph representation of information describing a print of the Rijksmuseum and an annotation acquired through Accurator is depicted in Figure~\ref{fig:data_model}. Constructs from the Europeana Data Model\footnote{\url{http://pro.europeana.eu/edm-documentation}} are used to model the metadata describing the object. An aggregation connects the metadata of the object with a digital representation of the object, in this case, an image. The identifier of the object is connected to metadata such as the title of the object and its subject matter. For this print, the subject matter is an Iconclass concept, representing owls. Contributors extend this existing information by adding annotations. New annotations are modeled according to the Web Annotation Data model\footnote{\url{https://www.w3.org/TR/annotation-model/}}, shown as the white ovals in the figure. One annotation has as a target the object, as well as an area of the digital representation of the object. Coordinates formalizing this area correspond to the bounding box drawn by the contributor. The body of the annotation corresponds to the value selected by the contributor, in this case, a concept from the IOC bird list, with the scientific name \textit{Bubo bubo}.

Using concepts instead of plain text to store annotations has a number of advantages. Concepts can have multiple labels, in different languages. The bird on the print of Figure~\ref{fig:data_model} can for example be identified by its scientific name \textit{Bubo bubo} as well as its common name in English, Eurasian eagle-owl. When a contributor enters one of these values, they refer to the same species concept. The common name in Dutch can now be used to retrieve the annotated object. This is a significant advantage over annotation using plain text since the annotations do not have to be translated every time a new language is supported. The hierarchy of some vocabularies has additional benefits, for annotation and subsequent object retrieval. More general concepts can be used during annotation at the moment a contributor cannot pinpoint a specific concept. During retrieval, the tree structure can be leveraged in the other direction, if someone searches for a general concept, more specific concepts lower in the hierarchy can be included in the results as well.

Contributors can explore the collection loaded using the semantic search functionality of the Accurator tool. The search is based on a graph search algorithm, which matches keyword queries with labels in the triple store. The graph structure is used to find connected objects and clusters similar objects together~\citep{Wielemaker2008}. Users can use this search functionality to explore the collection and find objects to annotate. Search thus functions alongside task assignment as an additional way of accessing tasks. The search algorithm is adapted to interpret added annotations as subject matter metadata, which allows users to directly inspect the result of their efforts. Observing that annotations improve the accessibility of the collection can be an added incentive to keep contributing. 

The Accurator tool provides multiple options to export data: annotations can be queried and exported to spreadsheets or RDF files. A public endpoint is available for queries and this can for example be used by systems that integrate multiple cultural heritage collections~\citep{Dijkshoorn2017a}. Additionally, the annotations are stored using a version management repository which can be easily published online, thereby making the results available to others.

\section{Case studies}
\label{sec:case_studies}
We validate the Accurator methodology using three real-world case studies in the form of nichesourcing campaigns. These show that the methodology is applicable in the highly different domains of birds on art, bible prints and fashion images. Table~\ref{tab:case_studies} provides a schematic overview of the cases, including links to online instances of the tool. In the following subsections, we describe each case in detail and discuss how the Accurator methodology and tool were implemented, listing the individual stages and steps of the methodology. In Section~\ref{sec:results}, we then provide an evaluation of quality and quantity of the resulting annotations.

\begin{table}
\scriptsize
\caption{Overview of the characteristics of the three case studies.}\label{tab:case_studies}
\centering
\begin{tabular}{| l | l | l | l |}
\hline

\textbf{Domain} & birds on art & bible prints & fashion images \\

\hline

\textbf{Goal of campaign} &
improve access &
support comparative research &
\begin{tabular}[t]{@{}l@{}} 
improve access\\ 
investigate vocabulary use
\end{tabular} \\

\hline

\textbf{Objects to annotate}&
\begin{tabular}[t]{@{}l@{}} 
2,160 artworks \tiny{\textit{(Rijksmuseum)}}\\ 
406 prints \tiny{\textit{(Naturalis)}}
\end{tabular}&
\begin{tabular}[t]{@{}l@{}} 
246 bible prints \tiny{\textit{(University Library)}}
\end{tabular}&
5,480 fashion objects \tiny{\textit{(Rijksmuseum)}}\\

\hline

\begin{tabular}[t]{@{}l@{}}
\textbf{Properties to annotate} \\ 
\textbf{\& candidate vocabularies}
\end{tabular} & 
\begin{tabular}[t]{@{}l@{}} 
33,799 taxons \tiny{\textit{(IOC bird list)}}\\ 
2 genders\\
3 stages of life\\
iconography
\end{tabular}&
\begin{tabular}[t]{@{}l@{}}
462 characters \tiny{\textit{(Bible ontology)}} \\
5,954 themes \tiny{\textit{(Iconclass)}} \\
34 emotions \tiny{\textit{(Emotion list)}} 
\end{tabular}&
\begin{tabular}[t]{@{}l@{}}
717 types \tiny{\textit{(Fashion thesaurus)}} \\
235 materials \tiny{\textit{(Fashion thesaurus)}} \\
117 techniques \tiny{\textit{(Fashion thesaurus)}} \\
20 colors \tiny{\textit{(Fashion thesaurus)}} \\
style period
\end{tabular} \\

\hline

\textbf{Niche community} & 14 birdwatchers & 7 bible experts & 18 fashion experts \\

\hline

\textbf{Tool} & \href{http://annotate.accurator.nl}{annotate.accurator.nl} & \href{http://bijbel.accurator.nl}{bijbel.accurator.nl} & \href{http://annotate.accurator.nl}{annotate.accurator.nl} \\

\hline

\textbf{Event} &
\begin{tabular}[t]{@{}l@{}}
birdwatching event\\
\tiny{\textit{(4-10-2015 Rijksmuseum)}}\\
\end{tabular} &
\begin{tabular}[t]{@{}l@{}}
bible event\\
\tiny{\textit{(4-4-2016 University Library)}}\\
\end{tabular} &
\begin{tabular}[t]{@{}l@{}}
stitch by stitch event\\
\tiny{\textit{(23-4-2016 Rijksmuseum)}}\\
\end{tabular} \\

\hline

\textbf{Quality assessment} & gold standard & review by professional & sample review by professionals \\

\hline

\end{tabular}
\end{table}

\subsection{Birds on art}
\label{sec:birds}
\begin{figure}
        \centering
                \includegraphics[width=0.5\textwidth]{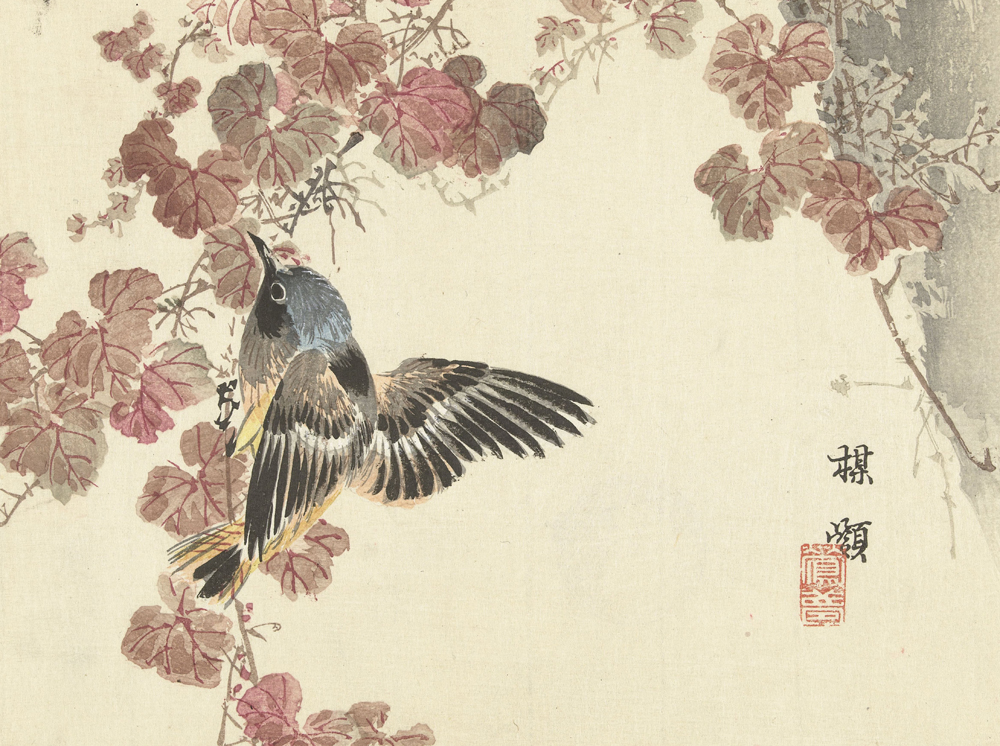}
        \caption{Print by Kono Bairei,  titled ``Bird and red vine''.}
        \label{fig:bird_print}
\end{figure}

The Rijksmuseum Amsterdam\footnote{\url{https://www.rijksmuseum.nl/en}} has a world-renowned art collection. The collection of works on paper is the largest sub-collection, although these prints, drawings and photographs are rarely on display due to their delicate nature. To improve access to the collection, the museum started to register, describe, digitize and publish the objects online~\citep{Dijkshoorn2012}. The subject matter of the prints is diverse and often outside the area of expertise of the employees, who mostly have an art-historical background. At times this results in overly general descriptions, such as the description of the Japanese print of Figure~\ref{fig:bird_print}: ``blue headed bird, near red vine''. Although this covers what is depicted, identifying the bird and tree would make the subject more specific.

\paragraph{Orientation}
In collaboration with the museum we involved experts in the process of accurately describing subject matter in order to improve access to the collection for online visitors \textit{(determine goal)}. The first type of subject matter that the museum tried to address regarded birds. The query functionality of the museum collection management system was used to define a set of artworks depicting birds \textit{(identify objects)}. In this case, existing descriptions served as a sufficient basis to identify 2,160 objects. The main goal of the campaign was to accurately identify the depicted species and add this to the objects' metadata \textit{(determine properties and select vocabularies)}. The IOC World Bird List, a comprehensive taxonomy of birds, was identified as candidate vocabulary. Other properties regarded the gender and age of the identified bird. The museum was interested whether the contributors could identify iconographic information related to the depicted birds as well.

Many birdwatchers go out into nature every week to seek birds. The museum identified them as the group of enthusiast that it was looking for \textit{(identify niche community)}. To be able to address potential contributors within the niche community, the Naturalis Biodiversity Center~\footnote{\url{http://www.naturalis.nl/en/}} was contacted. This natural history museum has access to many communities, including birdwatchers. Naturalis provided an additional set of 406 prints with realistic depictions of birds, which were already annotated by the head of the vertebrate collection and could serve as gold standard for evaluation purposes.

\paragraph{Implementation}
The Accurator annotation tool was deployed on a server and an export of metadata of the set of objects was loaded\footnote{\url{https://www.rijksmuseum.nl/en/api/rijksmuseum-oai-api-instructions-for-use}}, along with a conversion of the bird list\footnote{\url{https://github.com/rasvaan/ioc}} \textit{(deploy tool)}. The tagline and images were changed to suit the bird domain. Short instructions for the annotation fields were written and the bird list was related to the scientific name and common name fields \textit{(deploy tasks)}. A questionnaire inquiring about the experts' experience annotating artworks was created, to be handed out after the annotation session \textit{(setup user feedback elicitation)}. The annotation event was scheduled to coincide with World Animal Day, making the event easier to market \textit{(formulate marketing strategy)}. A page was created on the museum's website, advertising the event and annotation tool. The biodiversity center spread the invitation to appropriate channels and the event was picked up by national broadcasters.

\paragraph{Execution}
Two pilot events preceded the event, to test the stability of the system and to make employees of the biodiversity center and museum familiar with the system \textit{(run pilot)}. The two successful pilot events resulted in small incremental updates of the system, after which the organization of the main event could start. The birdwatching event was the first event organized as part of a nichesourcing campaign and set to take place in the historical library of the Rijksmuseum \textit{(organize events)}. To give experts an incentive to join the event, it was accompanied by various presentations related to the subject. After these talks, two and a half hours were spent annotating objects. The annotation session was closed by the curator of the museum after which people could join a bird-oriented guided tour through the museum. 

Fourteen birdwatchers annotated objects during the event. Many of them brought their own books of reference (in this case bird guides) and they often formed small groups, among which tasks were discussed. For many, this was a slow paced-process, annotations were thoroughly contemplated and values for all requested data were given when possible. A flyer explained how experts could use the system at home \textit{(run tasks online)}. After the event, no followup email was send to report on the results of the event or invite people to continue annotating, which we identified as a missed opportunity for advertising the online system.
 
\paragraph{Evaluation}
We used the gold standard of the Naturalis-provided prints to assess annotation quality \textit{(assess quality)}. It was not possible to feed back the results of the campaign into the collection management system of the museum, since adaptions had to be made to allow representing scientific species \textit{(measure impact)}. Comments on the functionality of the annotation system were collected during the event and using the questionnaire. The annotations and questionnaires were analyzed \textit{(analyze user feedback)} and in Section~\ref{sec:results} we discuss the results in more detail.

\subsection{Bible prints}
\label{sec:bibles}
Our second case study concerns 18th-Century picture bibles. In collaboration with historians and the university library of the Vrije Universiteit Amsterdam, a nichesourcing campaign was conducted to enable a comparison of bibles, belonging to the Dutch Protestant heritage collection of the library.

\begin{figure}
        \centering
                \includegraphics[width=0.5\textwidth]{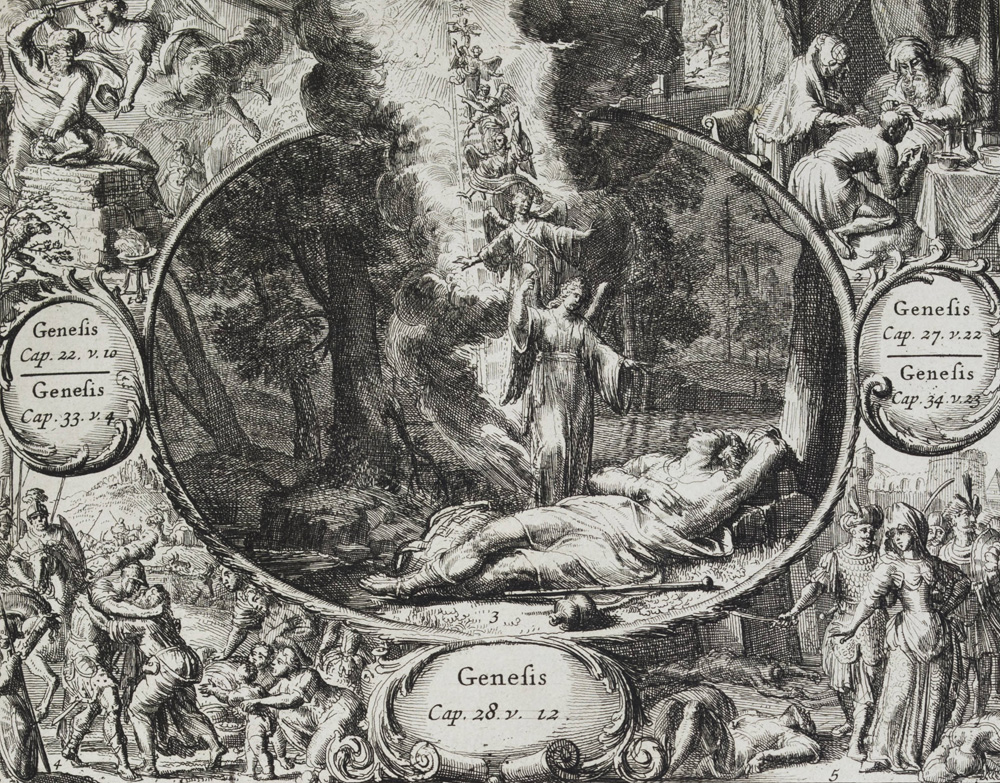}
        \caption{Print from Keur bible, depicting multiple biblical themes.}
        \label{fig:bible_print}
\end{figure}

\paragraph{Orientation}
A peculiar thing about picture bibles is that a buyer could commission which prints should accompany the religious texts~\citep{Stronks2011}. The prints depict bible scenes and were created by renowned artists. Figure~\ref{fig:bible_print} shows a digitized bible print from the collection. Analyzing which prints are included can shed light on aspects such as the popularity of artists as well as bible themes \textit{(determine goal)}. For the historians to be able to compare bibles, the pages and the prints among them had to be annotated. The historians selected two bibles that would be interesting to compare: one printed in 1728 by de Hondt and one printed in 1729 by the brothers Keur \textit{(identify objects)}. On request, pages of the two bibles were scanned by an external company, resulting in a total of 1,003 images.

The priority of the researchers and the university library was to gather data about the subject matter of prints \textit{(determine properties and select vocabularies)}. Two suitable structured vocabularies were identified for providing autocompletion alternatives: the bible ontology\footnote{\url{http://bibleontology.com}} is a source of biblical characters and the Iconclass vocabulary includes descriptions of many biblical themes. The historians were also interested in exploring changes in emotional expressivity depicted on the prints. To allow annotation of emotions a new vocabulary was created, based on a list of emotions of 18th-century theater texts, composed by the historians\footnote{\url{https://github.com/LaraHack/emotion_ontology}}.

For annotating the subject matter of bible prints, an expert has to be knowledgeable about bible scripture \textit{(identify niche community)}. The collaboration with the university library led to a fitting niche. The library regularly organizes seminars for ``friends of the university library'', which often revolved around biblical topics. Since these friends of the library were willing to attend events, the library anticipated that they might also be willing to join annotation events.

\paragraph{Implementation}
The Accurator annotation tool was installed on a university server and customized to accommodate the bible domain. Available metadata was exported from the library catalog\footnote{\url{https://github.com/VUAmsterdam-UniversityLibrary/ubvu_bibles}} and loaded in the annotation tool, together with the three candidate vocabularies \textit{(deploy tool)}. Tasks were defined by adding the fields biblical person, theme and emotion. These fields were related to parts of the candidate vocabularies and a description of the request \textit{(deploy tasks)}. The questionnaire used for the bird domain was adapted, now inquiring about the experience of annotating biblical themes, characters and emotions \textit{(setup user feedback elicitation)}. The library contacted the bible experts and dedicated seminars to annotation events \textit{(formulate marketing strategy)}.

\paragraph{Execution}
A pilot event was organized, during which two talks given by historians provided introductions to crowdsourcing and emotions in picture bibles \textit{(run pilot)}. This did not leave enough time for an extensive annotation session, although subsequent communications with the participants led to a number of observations. Annotating bible prints required elaborate instructions about the depth and thoroughness of requested annotations. Furthermore, we observed that bible experts not necessarily are experts in recognizing depicted 18th-century emotions. Additionally, many of the digitized pages where either blank or contained only text, making it nonsensical to ask experts to annotate subject matter.

The subset selection and information to gather was adapted based on the pilot event. The task of annotating emotions was removed, to be accomplished at a later time by some other niche community. The digitized pages were classified with whether the page depicts a biblical scene. This was another annotation task, but did not require expert knowledge and hence this task was accomplished using a regular crowdsourcing campaign. 246 pages depicted biblical themes and were included for the remainder of the nichesourcing campaign. In addition, a detailed step-by-step instruction manual was created to instruct people how to use the annotation tool.

For the main annotation event, the introduction was shortened, leaving more room for annotating prints \textit{(organize events)}. A computer room of the university library was used to host the event, with the addition of a hands-on experience with the two original historical picture bibles. Eight friends of the university library attended the annotation event and spent two and a half hours annotating bible prints. After the annotation events, participants were informed about the results of the annotation event and invited to further contribute using the online instance of the annotation tool \textit{(run tasks online)}.
  
\paragraph{Evaluation}
The annotations resulting from the events and from participants continuing at home were reviewed by library staff \textit{(assess quality)}. Verified annotations were exported from the annotation system, published and used by the library \textit{(measure impact)}. The library imported the annotations in its catalog, which currently allows browsing based on biblical characters and themes\footnote{This link for example lists all prints with a depiction of Moses: \url{http://imagebase.ubvu.vu.nl/cdm/search/collection/bis/searchterm/mozes/}}. Input for subsequent events was obtained during the event and from the questionnaires \textit{(analyze user feedback)}.

\subsection{Fashion images}
\label{sec:fashion}
In spring 2016, the Rijksmuseum organized an exhibition called Catwalk, during which fashion objects such as dresses and costumes were displayed\footnote{\url{https://www.rijksmuseum.nl/en/catwalk}}. The museum chose this well-advertised exhibition as the context for a nichesourcing campaign.

\begin{figure}
        \centering
                \includegraphics[width=0.5\textwidth]{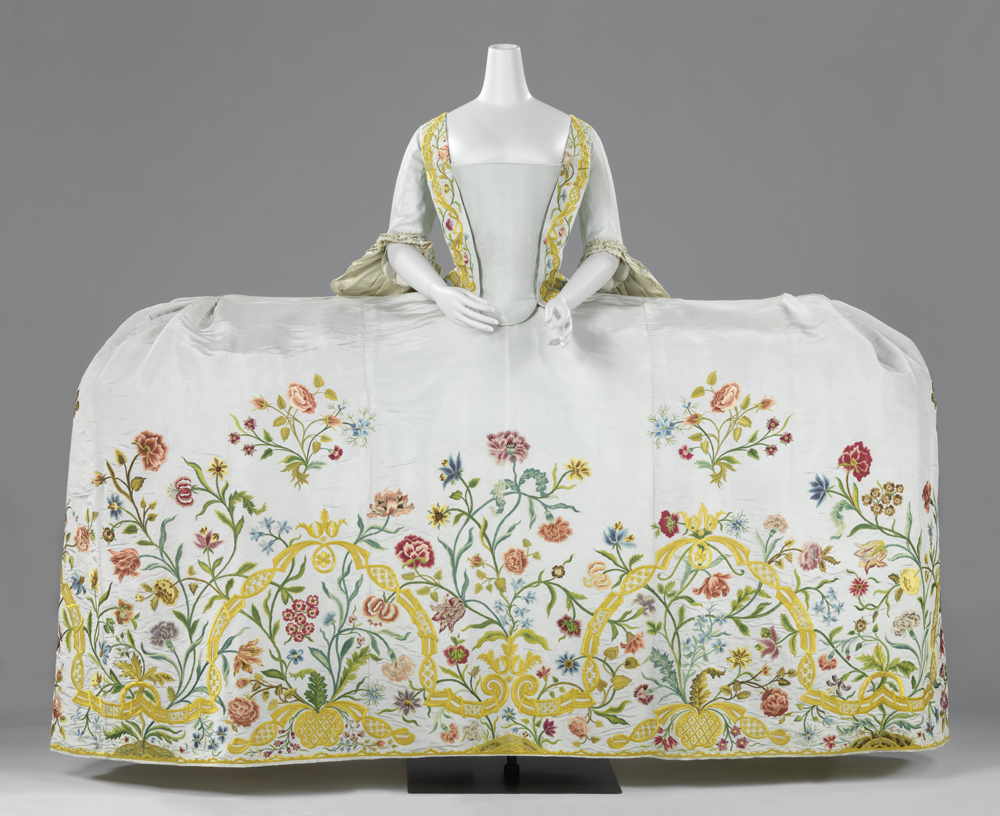}
        \caption{Dress with train, anonymous.}
        \label{fig:dress}
\end{figure}

\paragraph{Orientation}
The goal of the campaign was to better describe fashion objects, thereby improving online access \textit{(determine goal)}. Besides the museum, another user took interest in the annotations. A second goal was to support a researcher who develops a fashion thesaurus and wanted to investigate which terms contributors use to describe fashion objects. The types of objects in the fashion domain are more diverse than the prints and paintings from the previous two case studies. The museum owns a wide range of historical fashion objects, ranging from the dress depicted in Figure~\ref{fig:dress}, to jewelry and prints from fashion magazines. Since the domain included such diverse types of objects, multiple subsets were identified as relevant to the fashion domain, amounting to a total of 5,480 objects \textit{(identify objects)}. 

While the objects are diverse, an information specialist of the museum determined that the information that can be gathered about the objects can be categorized under general topics \textit{(determine properties and select vocabularies)}. These topics included technique, material, style period and color. A survey of structured vocabularies resulted in multiple possible candidates per topic, including the Art and Architecture Thesaurus (AAT) of the Getty\footnote{\url{http://www.getty.edu/research/tools/vocabularies/aat/}}. For the campaign it was however decided to use the fashion thesaurus created by Europeana\footnote{\url{http://skos.europeana.eu/api/collections/europeana:fashion.html}}, which is based on the AAT, but focusses more specifically on the fashion domain.

The diversity of the fashion domain makes it harder to pinpoint one niche community that is knowledgeable about all facets of the domain \textit{(identify niche community)}. The community of fashion enthusiast (fashionistas) is interested in fashion in a broader sense, but they might not know much about historical objects. There are many experts working with fashion on a professional level, but describing a shoe is something completely different from describing a lace detail. Therefore, to cover as much of the diverse fashion domain as possible, the museum had to turn to a more heterogenous group of people than the previous two case studies.

\paragraph{Implementation}
The collection data and structured vocabularies were loaded in the Accurator tool the Rijksmuseum already used for the birdwatching event \textit{(deploy tool)}. For the fashion domain six sub-domains were added: jewelry, accessories, fashion prints, paintings, costumes and lace. Contributors could choose one of these subdomains or the general fashion domain (which includes all objects of the sub-domains), to start adding annotations to. Similar tasks were deployed for each of these sub-domains, relating parts of the Europeana fashion thesaurus to requests for information regarding technique, material, style period and color \textit{(deploy tasks)}.

A questionnaire focussed on the fashion domain was created to elicit feedback \textit{(setup user feedback elicitation)}. The organization of ModeMuze\footnote{\url{https://www.modemuze.nl}} was willing to help address niche communities \textit{(formulate marketing strategy)}. ModeMuze is a Dutch aggregator of digitized fashion collections. This community was addressed and invited to participate in the event. Additionally, the Catwalk exhibition concluded with a conference for fashion professionals from the cultural heritage sector. An invitation was sent to these professionals as well. The event was organized following the conference, allowing professionals that attended the conference to join the annotation event. 

\paragraph{Execution}
Two fashion professionals participated in a small pilot, which did not bring major problems to light \textit{(run pilot)}. The main annotation event took place in the library of the museum \textit{(organize events)}. Since many of the contributors attended a conference in the days preceding the event, introductory talks were limited to an introduction of the annotation tool, leaving plenty of time to annotate objects. The broad invitation to different niche communities led to a diverse group of 18 contributors, including tailors, fashion curators and fashionistas. All of these contributors were asked to join in a discussion at the end of the event, discussing the campaign. The annotation tool stayed online following the event \textit{(run tasks online)}.

\paragraph{Evaluation}
To assess the annotation quality, three fashion professionals evaluated a sample of the collected annotations \textit{(assess quality)}. The free text annotations were compared with the structured vocabulary, seeing whether strong differences occurred, serving as input for the researcher interested in developing a fashion thesaurus for the cultural heritage domain \textit{(measure impact)}. Furthermore, from the discussion following the event and the questionnaires filled in during the event we received rich feedback on the annotation tool and what information about fashion can be collected \textit{(analyze user feedback)}. The results of the analysis of the questionnaires and the annotations are given in Section~\ref{sec:results}.

\section{Results}
\label{sec:results}
In this section, we discuss the results of the three nichesourcing campaigns. The quantity and quality of annotations provided by contributors are analyzed in Section~\ref{sec:annotations}, for each of the case studies. Section~\ref{sec:questionnaire} comprises the outcomes of a user evaluation of the Accurator annotation tool, which supported the campaigns.

\subsection{Analysis of annotations}
\label{sec:annotations}
For each case study, we analyze the annotations provided in terms of two dimensions: the number of provided annotations and the quality of the annotations. For the quantitative analysis, we split the numbers by annotation field. This provides an indication of whether a field was suitable for a domain. Furthermore, each of these fields is split according to the type of input, differentiating between text input and the input of concepts from vocabularies. This shows whether a vocabulary covered the values adequately. The last differentiating factor is the moment the annotation was entered. This is either during the event or during the subsequent possibility to add annotations online. This allows comparing the effectiveness of the campaigns of the three case studies. Regarding the qualitative analysis, for the bird case study a gold standard was available, allowing validation of the species annotations. The bible annotations and a subset of the fashion annotations were reviewed by professionals, providing an indication of their validity.

\begin{figure}
        \centering
                \includegraphics[height=7.5cm]{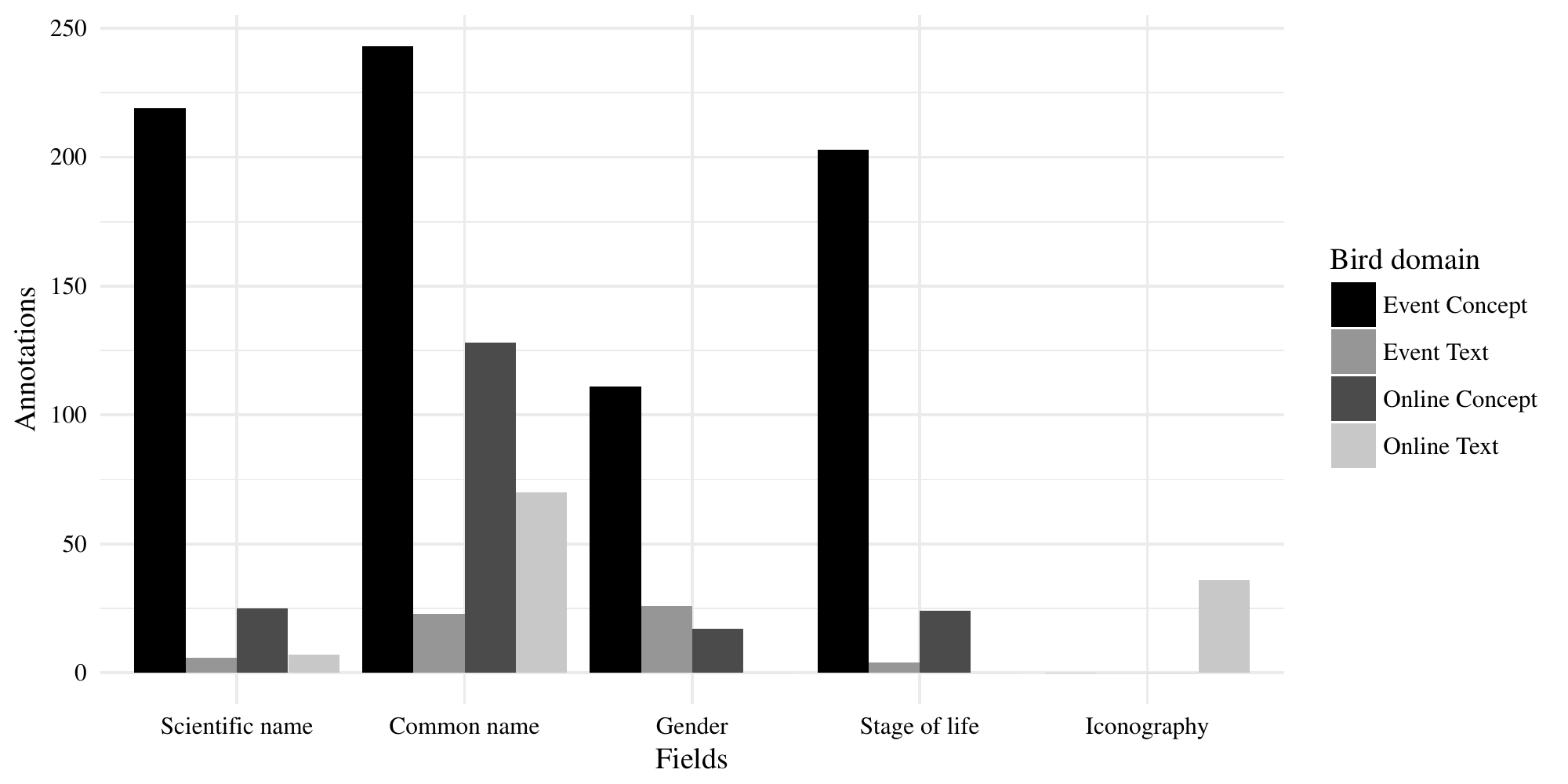}
        \caption{The number of annotations provided by contributors during the bird nichesourcing campaign, split by field, type of input and context of data entry.}
        \label{fig:bird_annotations}
\end{figure}

The birds nichesourcing campaign resulted in a total of 1,106 annotations, of which 835 annotations were entered during the event and 307 online. The contributors entered on average 59.6 annotations during the event. 65\% of the annotations concern species and 85\% of these 721 species annotation are concepts from the IOC bird list. During the annotation event slightly more common names (266) than scientific names (225) were entered. The opposite can be observed of the annotations entered online, here there are 198 common names and 32 scientific names entered. The iconography field is rarely used: During the event nothing was entered in this field, while online the field was used 36 times. The annotations provided during the event mainly concerned the Naturalis collection, which were not artistic interpretation of birds, hence the low count of iconography. A total of 231 stages of life annotations were added and 154 gender annotations. Concepts were used for the vast majority of these annotations.

The Naturalis prints allowed for evaluating the quality of the provided annotations, since the depicted species were already annotated by the head of the vertebrate collection. We compare the annotations entered by contributors to this gold standard and distinguish two types of matching. The first type is a direct match of the species concept provided by the professional and the annotation of the contributor. The second type of matching concerns concepts provided by the contributors that are one step higher up in the species taxonomy, thereby matching on a more generic level with the depicted species. Out of the 427 species annotations added to a print with gold standard, 344 of the annotations (80\%) exactly match with the annotation of the professional and 11 annotations (3\%)  match on a more general level.

\begin{figure}
        \centering
                \includegraphics[height=7.5cm]{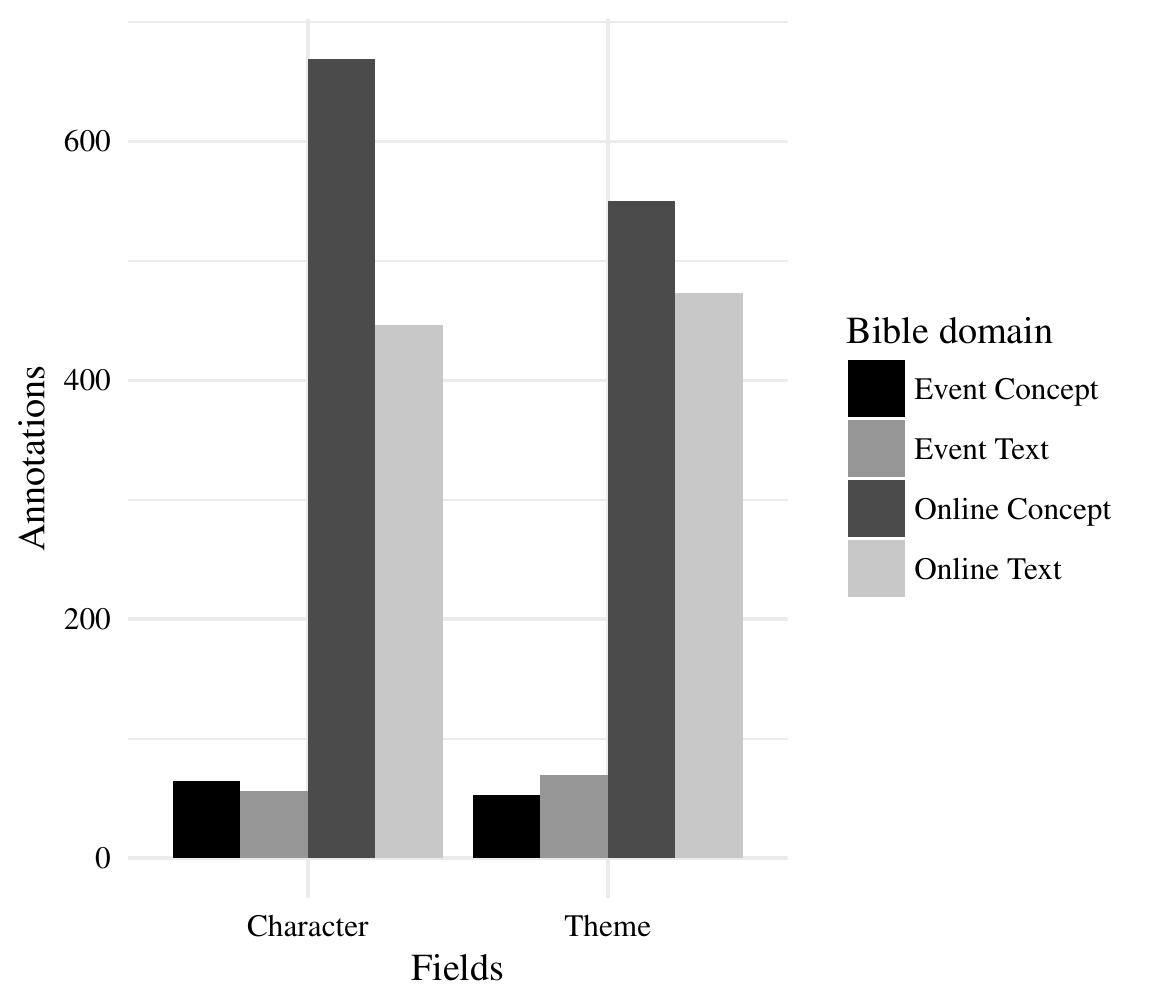}
        \caption{The number of annotations provided by contributors during the bible nichesourcing campaign, split by field, type of input and context of data entry.}
        \label{fig:bible_annotations}
\end{figure}

2,382 annotations were obtained during the bible campaign. An overview of the obtained annotations is given in Figure~\ref{fig:bible_annotations}. The event resulted in 244 annotations, a contributor added 34.9 annotations on average. In contrast to the other two domains, which have a low number of annotations added online, 90 percent of the bible annotations were obtained online. In total, 1,236 biblical characters were annotated, slightly more than the 1,146 themes. Vocabulary concepts were more often used than text annotations: 56 percent of the annotations. However, the use of concepts from structured vocabularies is lower than for example the species annotations within the bird domain. In July 2016, personnel of the university library reviewed all annotations available at that moment: 1,455 annotations in total. 96\% of the annotations were accepted: 630 theme and 764 bible character annotations. These verified annotations have been added to the libraries' catalog.

\begin{figure}
        \centering
                \includegraphics[height=7.5cm]{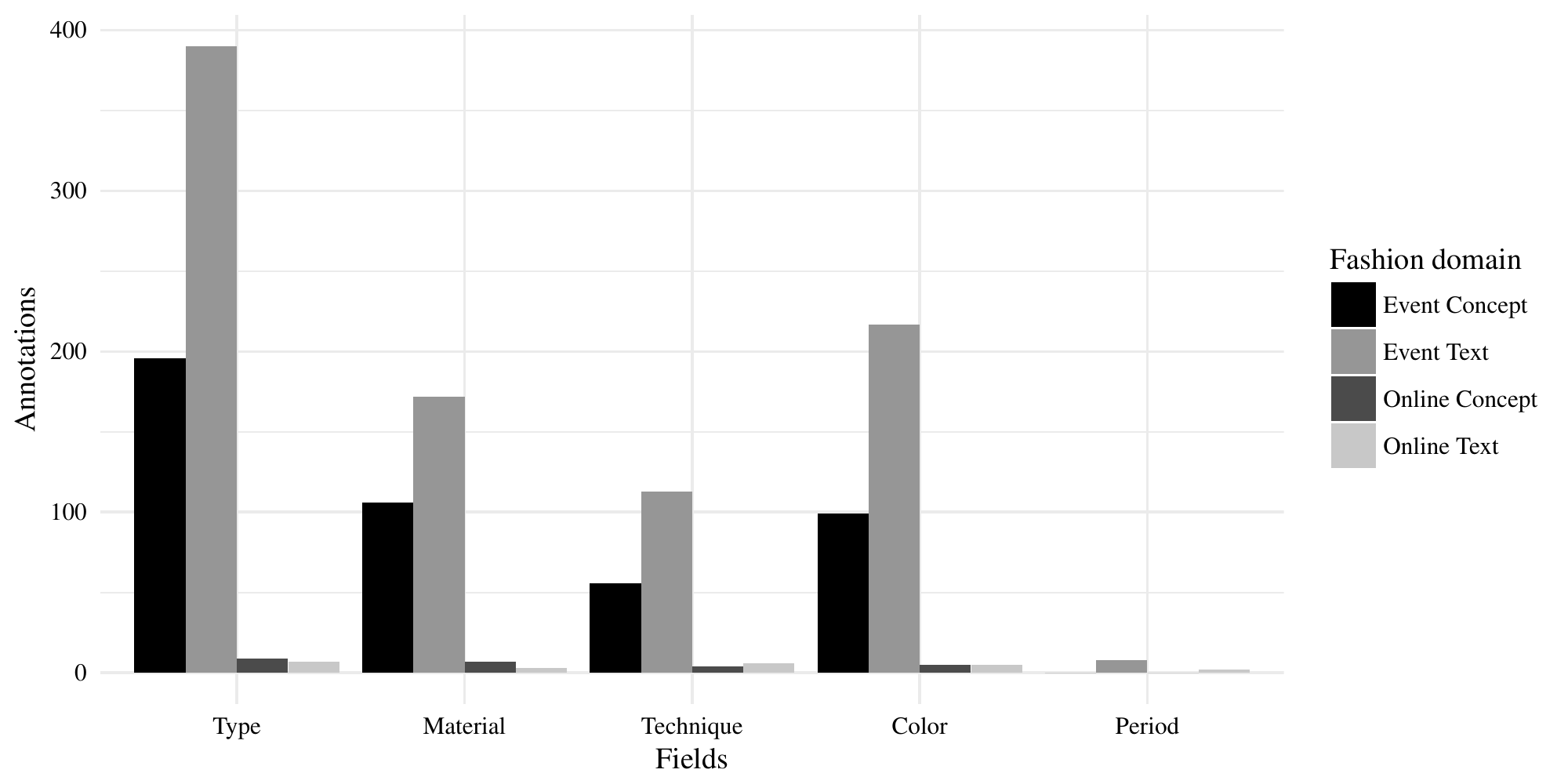}
        \caption{The number of annotations provided by contributors during the fashion nichesourcing campaign, split by field, type of input and context of data entry.}
        \label{fig:fashion_annotations}
\end{figure}
 
A total of 1405 annotations were added during the fashion campaign, as shown in Figure~\ref{fig:fashion_annotations}. Just 48 annotations were a result of the online campaign, 97\% of all annotations were a result of the event. During this event contributors added 75 annotations on average. 602 annotations concerned types of objects, 288 materials, 179 techniques and 326 colors. Style periods were rarely added, just 10 times. In contrast with the other two domains the use of concepts is low: 34\% of the total annotations originate out of the Europeana Fashion Thesaurus, the rest are textual annotations.

A sample of 40 annotations was evaluated by 3 fashion professionals. One professional works at the Rijksmuseum, one works for the fashion aggregator Modemuze and the last for a fashion museum in Antwerpen. Ten annotations were randomly picked from respectively the type, material, technique and color annotations. For each annotation the professionals were asked whether it was correct, incorrect, or whether they were unable to assess it. We used majority voting to reach an assessment for 37 of the annotations, for 3 annotations the evaluations were inconclusive. Out of the sample, 89\% of type annotations, 78\% of the material annotation, 78\% of the technique annotations and 90\% of the color annotations were deemed to be correct. From the 37 annotations upon which agreement was reached, 84\% were considered correct.

\subsection{Evaluation annotation tool}
\label{sec:questionnaire}
In order to evaluate effectiveness of the Accurator annotation tool, at the end of each of the annotation events questionnaires were handed out. 14 birdwatchers, 9 bible experts and 18 fashion experts filled in the questionnaire. In this section we list the outcomes, focussing on the discussion around task assignment and usability of the Accurator annotation tool. 

During the three campaigns, different settings for task assignment were used, which are described in Section~\ref{sec:task_assignment}. The bible domain used the ranked setting, the fashion domain the sub-domain based setting and the bird domain used recommendation. Since the latter two are more advanced ways of assigning tasks to contributors, questions of how these settings were experienced by the contributors were included. The sub-domain based setting of the fashion domain is deemed useful by 89\% of the respondents. Contributors comment that using the sub-domains it is easier to access objects they know something about. Some would like the option to refine a sub-domain by adding filters, thereby, for example indicating the type of accessories recommended. Additionally, dividing the domains based on style period would be appreciated. 79\% of the bird watchers finds recommendation based on expertise useful. The comment that it makes the process more efficient, although a different elicitation of expertise is proposed by many contributors. Expertise topics concerned different families of the taxonomy, while many think it would be more useful to ask how much someone knows about a certain region where the birds occur.

\begin{figure}
        \centering
                \includegraphics[width=1\textwidth]{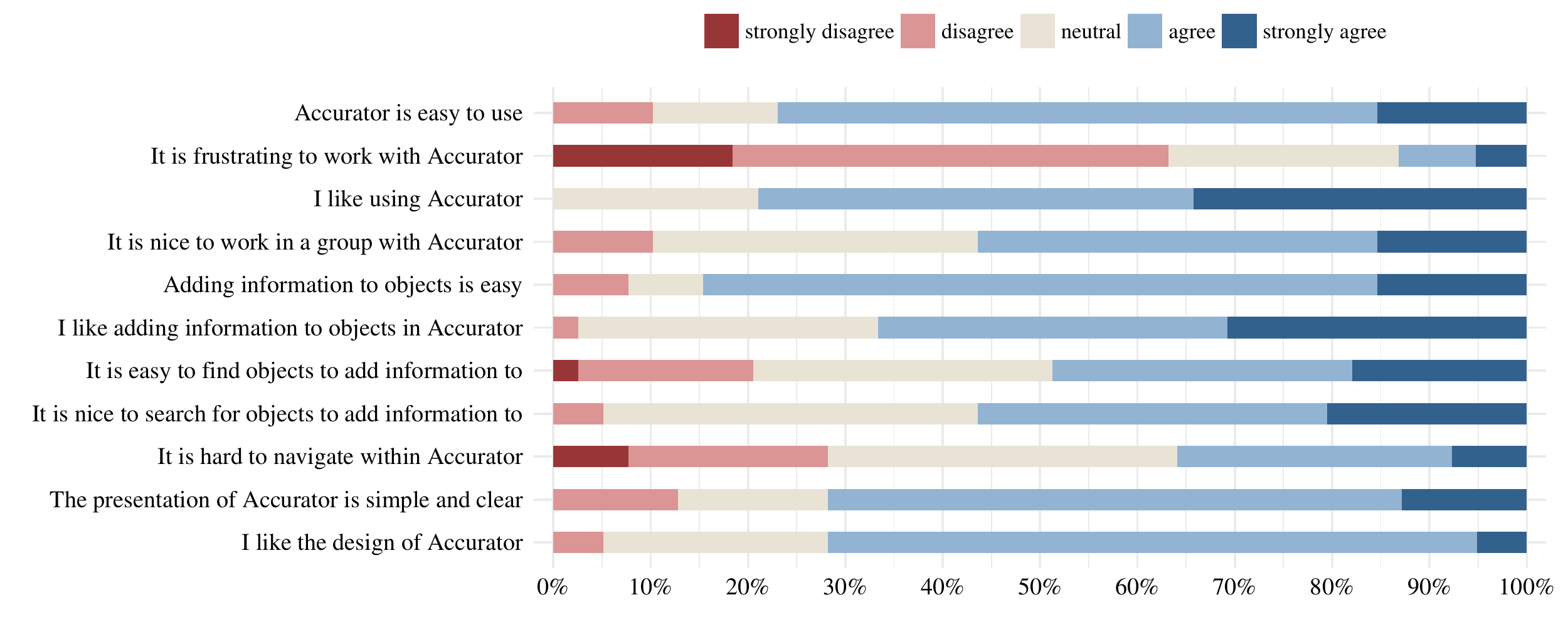}
        \caption{Overview of the answers provided that regard the usability of the Accurator annotation tool.}
        \label{fig:usability}
\end{figure}

The questionnaire included 11 statements regarding the usability of the annotation tool. Participants were asked to indicate their agreement on a five-level Likert scale, ranging form ``strongly disagree'' to ``strongly agree''. Figure~\ref{fig:usability} shows an overview of the answers of the contributors of the three case studies combined, which sums up to a total of 41 questionnaires. The evaluation of the general usability of the tool is good, 77\% of the contributors agree that Accurator is easy to use. 63\% disagrees with that it is frustrating to use the tool. An even higher number of contributors (79\%) liked using Accurator, no one disagreed on this point. Over half of the respondents enjoyed working with the tool in a group, although an additional one-third of the responses regards this point as neutral.

The questionnaires also included more focused questions, regarding the ability to add information to objects, searching for objects and the design of the tool. 85\% of the respondents finds it easy to add information to objects, two-thirds of the respondents like adding information to objects using the tool. Navigating to objects is deemed more complicated, over half of the respondents disagreed or replied neutrally on the question whether it was easy to find an object to add information to. Over half of the contributors (56\%) did enjoy searching for objects. Navigating the tool is found to be more complicated, 36\% agreed that it is hard to navigate Accurator. Over two-thirds of the contributors like the design of Accurator and find the tool simple and clear.


\section{Discussion \& Future Work}
\label{sec:discussion}
Nichesourcing is a method for outsourcing tasks that require a significant level of expertise in a specific domain. The methodology presented in this paper is geared towards executing challenging annotation tasks in the cultural heritage domain in a sustainable and repeatable fashion. The annotation events are central to the methodology and the enthusiasm with which people shared their knowledge and showed the potential of this method. The Accurator image annotation tool supports the methodology and a user evaluation indicates that the design and usability of the tool are appreciated, as well as working together with other members of the community. The three case studies show that the nichesourcing methodology in combination with the image annotation tool can be used to collect high quality annotations in a variety of domains and annotation tasks.
  
While all three case studies required experts to be knowledgable about the domain on hand, annotating fashion images proved to be most challenging. Determining materials and techniques from single images is difficult and the formulated requests for annotations proved to be more ambiguous as well. Furthermore, the use of terms from structured vocabularies differs significantly per case study. The amount of concepts used is an indicator of how suitable a vocabulary is to describe a property of a collection object. The difference in collected annotations between the event and online tool underlines the importance of a strong marketing strategy. After the bible annotation event multiple e-mails were sent inviting people to keep contributing, which clearly shows in the results.

The methodology outlined in this paper is also applicable outside the cultural heritage domains. At the moment annotation tasks require expert-knowledge and niche communities with that knowledge can be identified, the nichesourcing methodology can be used. Tasks can range from identifying species on camera trap images collected by biologists, recognizing musical instruments in audio recording, to identifying different types of cinematography in videos. The Accurator annotation tool is just one example of the tools that can be used to collect annotations, this time focused on images. The tool deployed in the methodology can be replaced by other tools when needed, which could for example be suited to annotate sound or video clips.

In future campaigns we plan to optimize settings that impact results, such as the number of selected objects, the formulation of information requests and the influence of the marketing schedule. We plan to translate the social aspect of the annotation events into functionality of the tool and investigate whether this will retain more contributors. To accomplish goals more efficiently, we will investigate embedding nichesourcing in hybrid crowdsourcing workflows, splitting a campaign into subtasks that are solved using different methods within the human computation spectrum. This would have the benefit that for simple tasks that can be solved by anyone in the crowd we can resort to methods other than nichesourcing, thereby not wasting the goodwill of our expert volunteers on simple tasks. Other possibilities are automating parts of the campaign, such as utilizing computer vision to recognize objects on images. Finding a hybrid approach that strikes the right balance of quality and quantity of annotations will improve the usefulness of cultural heritage data published online.

\bibliography{bibliography}

\end{document}